**Delay estimation in a two-node acyclic network**

Author for Correspondence:

Radhakrishnan Nagarajan
Department of Biostatistics, Room: 3234
University of Arkansas for Medical Sciences
4301 West Markham, Slot 781
Little Rock, AR 72205
Email: nagarajanradhakrish@uams.edu

**Abstract**

Linear measures such as cross-correlation have been used successfully to determine time delays from the given processes. Such an analysis often precedes identifying possible causal relationships between the observed processes. The present study investigates the impact of a positively correlated driver whose correlation function decreases monotonically with lag on the delay estimation in a two-node acyclic network with one and two-delays. It is shown that cross-correlation analysis of the given processes can result in spurious identification of multiple delays between the driver and the dependent processes. Subsequently, delay estimation of increment process as opposed to the original process under certain implicit constraints is explored. Short-range and long-range correlated driver processes along with those of their coarse-grained counterparts are considered.

**1. Introduction**

Estimating delays from the observed processes has been an area of great interest both from theoretical and experimental standpoints. Inferring delays from temporal processes is an inverse problem and can also be useful in inferring causal relationships between them [1-3]. The present study investigates a primitive two-node acyclic network comprising of a driver and a dependent process with single and two delays, Fig. 1. We consider the class of drivers ($x$) whose auto-correlation functions $R_x(k) = E(x_n x_{n+k})$ are positive and decreases monotonically as a function of lag ($k$).

Classical delay estimation techniques using linear measures such as cross-correlation function are especially useful when the driver process is uncorrelated. The procedure begins, by estimating the cross-correlation functions $R_{xy}(k) = E(x_n y_{n+k})$ between the driver ($x$) and the dependent processes ($y$) as function of lag ($k$), Fig. 1. A significant non-zero cross-correlation at a given lag is chosen as the desired delay between $x$ and $y$. However, drivers need not necessarily be uncorrelated. A classic example is that of genetic networks [4, 5], where an up-stream gene (driver) with auto-regulatory feedback regulates a down-stream gene (dependent) through multiple pathways with distinct delays. In such cases, direct estimation of the delay between $x$ and $y$ from their observed values using measures such as cross-correlation may not be sufficient. Encouraged by such examples, we explore delay estimation from the *increment processes* as opposed to that of the original processes.

## 2. Methods and Results

## A. Statistically significant delays

In the present study, only positive cross-correlation estimates between the driver and the dependent processes are assessed for statistical significance. The cross-correlation estimate at a given lag is deemed significant if its value is considerably higher than those obtained on the random shuffled counterparts. A brief description of the procedure is enclosed below.

**Step 1** Estimate the cross-correlations as a function of the lags $R_{xy}(k), \tau = 1...T$ between the driver and dependent processes $x_n$ and $y_n$.

**Step 2** Generate random shuffled counterparts $x_i^*$ and $y_i^*, i = 1...n_s$ of $x_n$ and $y_n$ by resampling without replacement [6-8]. Estimate the cross-correlation as a function of the delays on the $n_s$ shuffled counterparts $R_{x_i^* y_i^*}^*(\tau), \ \tau = 1...\tau_{max}, i = 1...n_s$.

**Step 3** Cross-correlation estimate at lag $k$ is statistically significant if $R_{xy}(k) > R_{x_i^* y_i^*}^*(k), \forall i = 1...n_s$. This lag ($k$) is the desired delay between the driver and the dependent processes. Thus a one-side test is sufficient. The number of surrogates was fixed at $n_s = 99$, this corresponds to a significance level of $\alpha^+ = 1/(99+1) = 0.01$ [3-4] for a one-sided test [6-8].

In order to estimate statistically significant delays from the increment processes $\delta x_n = x_{n+1} - x_n$ and $\delta y_n = y_{n+1} - y_n$ repeat Steps 1, 2 and 3 for the increment processes.

Prior to a detailed discussion, the motivation behind the choice of delay estimation on the increment process is illustrated with a simple example.

**Example**: Consider a two-node acyclic network with a single delay

Driver ($x$) $\qquad\qquad$ : $x_1 x_2 x_3 \ldots\ldots x_n x_{n+1} x_{n+2} \ldots\ldots\ldots\ldots\ldots\ldots\ldots\ldots\ldots\ldots x_N$

Dependent ($y_n = x_{n-\tau}$) : $y_1 y_2 y_3 \ldots\ldots y_n y_{n+1} y_{n+2} \ldots\ldots y_{n+\tau} y_{n+\tau+1} y_{n+\tau+2} \ldots\ldots y_N$

White noise process ($e$): $e_1 e_2 e_3 \ldots\ldots e_n e_{n+1} e_{n+2} \ldots\ldots\ldots\ldots\ldots\ldots\ldots\ldots\ldots e_N$

Case (i) Uncorrelated Driver

*Delay estimation from the given processes*

Consider the uncorrelated driver ($x$) sampled from a white-noise process ($e$) with zero-mean and variance $R_e(0)$ (i.e. $x_n = e_n, n = 1\ldots N$) and the dependent process ($y_n = x_{n-\tau}$).

Cross-correlation estimates as a function of lag $k$ yields

$$E(x_{n+k}.y_{n+\tau}) = E(x_{n+k}.x_{n+\tau}) = E(e_{n+k}.e_{n+\tau}) = R_e(0), \text{for } k = 0$$
$$= 0, \text{for } k \neq 0 \qquad \ldots\ldots\ldots\ldots\ldots\ldots\ldots\ldots (1)$$

A positive cross-correlation estimate exists only for $k = 0$, which corresponds to delay $\tau$ between $x_n$ and $y_n$.

*Delay estimation from the increment processes*

Consider the increment processes $\delta x_n = x_{n+1} - x_n$ and $\delta y_n = y_{n+1} - y_n$.

$$E(\delta x_{n+k}.\delta y_{n+\tau}) = E(\delta x_{n+k}.\delta x_{n+\tau}) = E(e_{n+k}.e_{n+\tau}) = 2R_e(0), \text{for } k = 0$$
$$= 2R_e(k) - R_e(k+1) - R_e(k-1), \text{for } k \neq 0$$

$$\ldots\ldots\ldots\ldots\ldots\ldots\ldots\ldots. (2)$$

Unlike (1) cross-correlation of the increment processes persist for delays $k$ = -1, 0, 1. However, from (2) cross-correlation estimate is positive only for $k$ = 0 and negative for $k$ = -1 and 1. From our definition of statistical significance (Sec. A), cross-correlation estimate only at $k$ = 0 is statistically significant.

As a remark, it should be noted that it is possible to identify the delay even for nonlinearly correlated drivers ($x$) with fast *linear* de-correlation time comparable to that of white noise process using linear measures such as cross-correlation. An example of such a driver is a chaotic logistic map given by the expression $x_{n+1} = 4.x_n.(1 - x_n)$. Therefore, in the subsequent discussion the term *correlated drivers* implicitly refers to drivers whose *linear* de-correlation time is comparatively larger than those of white noise.

## Case (ii) Correlated Driver

Consider a driver process generated as linear combination of samples from a white noise process i.e. $x_i = e_i + e_{i+1}$, $i = 1...n - 1$. Let the dependent process be $y_n = x_{n-\tau}$.

*Delay Estimation from the given processes*

$$E(x_n.y_{n+\tau}) \ = R_x(0) = 2R_e(0) > 0$$

$$E(x_{n+1}.y_{n+\tau}) = R_e(0) > 0 \ldots\ldots\ldots\ldots\ldots\ldots\ldots\ldots\ldots\ldots\ldots\ldots\ldots\ldots\ldots. \ (3)$$

For the correlated driver, positive cross-correlation (3) persists for delays other than $\tau$. Such correlations are an outcome of the correlated nature of the driver and shall be referred to as *correlation leak* in the subsequent sections. Correlation leak can be statistically significant (Sec. A), and may imply spurious existence of multiple delays between the driver and the dependent processes.

*Delay Estimation from the increment processes*

$$E(\delta x_n . \delta y_{n+\tau}) = 2[R_x(0) - R_x(1)] = 2R_e(0) > 0$$

$$E(\delta x_{n+1} . \delta y_{n+\tau}) = 2.R_e(k) - R_e(k+2) - R_e(k-2) \text{ for } k \neq 0 \dots\dots\dots\dots\dots\dots\dots.. (4)$$

Unlike (3), cross-correlation analysis of the increment series (4) reveals statistically significant positive correlation (Sec. A), only at $k = 0$. While cross-correlation persists for lags $k$ = -2, 2, these are negative and not significant (Sec. A). Thus delay estimation on the increment process can be useful in minimizing contributions due to correlation leak.

Inspired by the above example, cross-correlation analysis of increment processes in conjunction with those of the original process in delay estimation in a two-node acyclic network is explored. As noted earlier, the driver processes is implicitly assumed to be positively correlated with monotonic decreasing auto-correlation function. In this respect, we discuss the results for short-range correlated stationary first-order Gauss-Markov driver process and long-range correlated stationary fractional auto-regressive integrated moving average driver (FARIMA) process [9, 10]. Instances of delay estimation on the coarse-grained counterparts of their increment series are also discussed.

## B. Short-range correlated driver

Short-range correlated stationary first-order Gauss-Markov process is given by the expression

$$x_t = \alpha x_{t-1} + e_t \dots\dots\dots\dots\dots\dots\dots\dots\dots\dots\dots\dots\dots\dots \ (5)$$

where $e_t$ sampled from a normally distributed white noise with zero-mean and unit variance. Since we consider positively correlated driver process with monotonic decreasing auto-correlation function we consider only processes (5) where $0 < \alpha < 1$. The corresponding auto-correlation function $R_x(k)$ for $0 < \alpha < 1$

$$R_x(k) = E(x_n x_{n+k}) = \alpha^k R(0) > 0 \forall k \dots\dots\dots\dots\dots\dots\dots\dots \ (6)$$

## B1. Short-range correlated driver and single-delay

An example of two-node acyclic network with single delay is shown in Fig. 1a. Consider the cases where the dependent node $y_n^o$ which lags the driver node by a delay $\tau$ given by

$$y_n^o = \beta x_{n-\tau} \text{ such that } \beta > 0, \ \tau > 0 \ \dots\dots\dots\dots\dots\dots\dots\dots\dots\. \ (7)$$

In (7), $\beta$ contributes to the overall variance of the process, hence can be factored out to obtain the normal form $y_n$,

$$y_n = \frac{y_n^o}{\beta} = x_{n-\tau} \dots\dots\dots\dots\dots\dots\dots\dots\dots\dots\dots\dots\dots\dots\dots\dots \ (8)$$

*Delay estimation from the original process*

Cross-correlation function between the driver $x_n$ (7) and the dependent $y_n$ (8) processes at lag $k$ is given by

$$R_{xy}(\tau - k) = E(x_{n+k} y_{n+\tau}) = E(x_{n+k} x_n) = R_x(k) \ \dots\dots\dots\dots\dots\dots\dots\.. \ (9)$$

Substituting for the auto-correlation function $R_x(\ )$ from (6) we get

$$R_{xy}(\tau - k) = \alpha^k R_x(0) > 0 \forall k \ \dots\dots\dots\dots\dots\dots\dots\dots\dots\dots\.... \ (10)$$

Since $0 < \alpha < 1$, we have $R_{xy}(\tau - k) < R_{xy}(\tau), \forall k > 0$. Therefore, irrespective of the choice of the process parameter $\alpha$, the driver and the dependent nodes are *maximally correlated* at lag $k = \tau$, which corresponds to the delay between $x_n$ and $y_n$. However, there is considerable positive *correlation leak* across lags $(\tau - k), k \neq 0$, whose magnitude $R_{xy}(\tau - k), k \neq 0$ increases as a function of the process parameter $\alpha$. The correlation leak is especially significant in the limit $\alpha \to 1$. An instance of cross-correlation estimate as a function of lags for the driver and the dependent processes (5 and 8) with parameters ($\alpha = 0.9$, $\tau = 10$) is shown in Fig. 2a. Statistically significant cross-correlation (Sec. A) is observed at a number of lags in addition to $k = \tau$. This is not a drawback of the estimation procedure but an inherent feature due to the correlated nature of the driver. As noted earlier (3), in such cases, it is possible to infer spurious existence of multiple delays (directional paths) from the driver to the dependent process.

*Delay estimation from the increment process*

Consider the increment processes $\delta y_{n+1} = y_{n+1} - y_n$ and $\delta x_{n+1} = x_{n+1} - x_n$ corresponding to $x_n$ (5) and $y_n$ (8). The corresponding cross-correlation function at lag $k$ is given by

$$R_{\delta x \delta y}(\tau - k) = E(\delta x_{n+k} \delta y_{n+\tau}) = 2R_x(k) - R_x(k+1) - R_x(k-1) \ \ldots\ldots\ldots\ldots\ldots\ldots (11)$$

Substituting for the auto-correlation function $R_x()$ from (6) we get

$$R_{\delta x \delta y}(\tau) = 2.(1-\alpha)R_x(0) > 0 \ \ldots\ldots\ldots\ldots\ldots\ldots\ldots\ldots\ldots\ldots\ldots\ldots\ldots\ldots\ldots.. (12)$$

$$R_{\delta x \delta y}(\tau - k) = -\alpha^{(k-1)}(1-\alpha^2) < 0 \ \ldots\ldots..\ldots\ldots\ldots\ldots\ldots..\ldots\ldots\ldots... (13)$$

From (12 and 13) we note that $R_{\delta x \delta y}(\tau - k) < R_{\delta x \delta y}(\tau)$. More importantly, we note that $R_{\delta x \delta y}(\tau) > 0$ whereas $R_{\delta x \delta y}(\tau - k) < 0, \forall k \neq 0$. An instance of cross-correlation estimates as a function of lags for the increment of the driver and the dependent processes with ($\alpha$ = 0.9, $\tau$ = 10) is shown in Fig. 2b. The cross-correlation estimate was statistically significant (Sec. A), only at lag $k = \tau$ which corresponds to the delay between the driver and the dependent processes. These results have to be contrasted with those of Fig. 2a, where the correlation leak $R_{xy}(\tau - k)$ resulted in identifying multiple delays between the driver and the dependent processes.

**Summary I** *For a two-node acyclic network with a single delay and Gauss-Markov driver with parameter $0 < \alpha < 1$, delay estimation using cross-correlation on the original process can result in significant positive correlation at several delays in addition to that of $\tau$, attributed to inherent correlation leak. These in turn may indicate spurious existence of multiple delays (directional paths) between the driver and the dependent processes. However, analysis on the increment processes resulted in positive cross-correlation only at lag corresponding to the delay between the driver and the dependent processes.*

## B2. Short-range correlated driver and two-delays

An example of two-node acyclic network with two delays and a correlated driver (5) is shown in Fig. 1b. The dependent process is generated as a linear combination of the driver (5) with delays $\tau_1$ and $\tau_2$ as

$$y_n^o = \beta_1 x_{n-\tau_1} + \beta_2 x_{n-\tau_2} \text{ such that } \beta_2 > \beta_1 > 0, \ \tau_2 > \tau_1 > 0 \ \dots\dots\dots\dots\dots\dots (14)$$

In order to obtain the normal form $y_n$ of $y_n^o$ we follow the steps below

$$y_n^o = \beta_2 \left( \frac{\beta_1}{\beta_2} x_{n-\tau_1} + x_{n-\tau 2} \right)$$

Substituting, $\beta = \frac{\beta_1}{\beta_2}$ such that $0 < \beta < 1$ in the above expression, we get

$$y_n^o = \beta_2 (\beta x_{n-\tau_1} + x_{n-\tau 2}) \qquad \text{I}$$

$$y_n = \frac{y_n^o}{\beta_2} = \beta x_{n-\tau_1} + x_{n-\tau 2} \text{ such that } \tau_2 > \tau_1 > 0,\ 0 < \beta < 1 \ldots\ldots\ldots\ldots\ldots\ldots\ (15)$$

In (15) $\beta_2$ affects the overall variance of $y_n^o$, hence can be factored out. In the subsequent discussion we shall only consider the normal form $y_n$ (15).

*Delay estimation from the original process*

From (15) we have

$$y_{n+\tau_1} = \beta x_n + x_{n-\tau}, \text{ where } \tau = \tau_2 - \tau_1 > 0 \ldots\ldots\ldots\ldots\ldots\ldots\ldots\ldots\ldots\ldots\ (16)$$

$$y_{n+\tau_2} = \beta x_{n+\tau} + x_n, \text{ where } \tau = \tau_2 - \tau_1 > 0 \ldots\ldots\ldots\ldots\ldots\ldots\ldots\ldots\ldots\ldots\ (17)$$

Their corresponding cross-correlation functions with $x_n$ (5) is given by

$$R_{xy}(\tau_1) = E(x_n y_{n+\tau_1}) = \beta.R_x(0) + R_x(\tau) > 0, \text{ where } \tau = \tau_2 - \tau_1 \ldots\ldots\ldots\ldots\ldots\ (18)$$

$$R_{xy}(\tau_2) = E(x_n y_{n+\tau_2}) = \beta.R_x(\tau) + R_x(0) > 0, \text{ where } \tau = \tau_2 - \tau_1 \ldots\ldots\ldots\ldots\ldots\ (19)$$

From (18 and 19) it can be seen that the magnitude of the cross-correlation between the driver and the dependent process is proportional to parameter $\beta$.

**Remark 1** $R_{xy}(\tau_2) > R_{xy}(\tau_1)$ …………………………………………............ (20)

Subtracting (18) from (19) we get

$$R_{xy}(\tau_2) - R_{xy}(\tau_1) = (1-\beta).[R_x(0) - R_x(\tau)]$$

Since $R_x(m) > R_x(n)$ for $m < n$ and $0 < \beta < 1$, $R_{xy}(\tau_2) > R_{xy}(\tau_1)$.

In the case of uncorrelated driver, the following inequality holds $R_{xy}(\tau_2) > R_{xy}(\tau_1) > R_{xy}(k) = 0$ for $k \neq \tau_1, \tau_2$. Thus ranking the cross-correlation function in descending order is useful in inferring the delays between the driver (5) and the dependent (15) processes. However, such a ranking need not necessarily hold in the case of correlated drivers. As correlation leak around delay $\tau_2$ can be significantly higher than that of $R_{xy}(\tau_1)$. This in turn implies that ranking the cross-correlation can result in spurious identification of delays between the driver and the dependent processes. In the following Remark, we derive a constraint on the process parameters ($\alpha$ and $\beta$) in order to preserve the ranking $R_{xy}(\tau_2) > R_{xy}(\tau_1) > R_{xy}(k)$ for $k \neq \tau_1, \tau_2$.

**Remark 2** *Constraint on the parameters $\alpha$ and $\beta$ such that $R_{xy}(\tau_2) > R_{xy}(\tau_1) > R_{xy}(k)$ for $k \neq \tau_1, \tau_2$.*

From (16), we have

$$E(x_{n+1} y_{n+\tau_2}) = R_{xy}(\tau_2 - 1) = E(x_{n+1} y_{n+\tau_2}) = \beta.R_{xy}(\tau-1) + R_x(1) \; …………..….… (21)$$

In order for the ranking to be preserved we need

$$R_{xy}(\tau_1) > R_{xy}(\tau_2 - 1) \; …………………………………………………..…… (22)$$

Since $R_{xy}(\tau_2) > R_{xy}(\tau_2 - 1)$ and $R_{xy}(\tau_2) > R_{xy}(\tau_1)$ from (20)

Substituting from (18) and (21) in (22) we get

$$\beta.R_x(0) + R_x(\tau) > \beta.R_x(\tau - 1) + R_x(1)$$

$$\beta > \frac{R_x(1) - R_x(\tau)}{(R_x(0) - R_x(\tau - 1))}$$

Substituting for the auto-correlation function $R_x()$ from (6) we get

$$\beta > \frac{\alpha - \alpha^\tau}{1 - \alpha^{\tau - 1}} = \alpha \ \text{ i.e. } \beta > \alpha \ \ldots\ldots\ldots\ldots\ldots\ldots\ldots\ldots\ldots\ldots\ldots\ldots.. \ (23)$$

Thus the constraint on the parameters ($\alpha$ and $\beta$) so as to preserve the ranking $R_{xy}(\tau_2) > R_{xy}(\tau_1) > R_{xy}(k)$ is $\beta > \alpha$.

Cross-correlation estimates $R_{xy}(k)$ between the driver and the dependent processes as a function of lag ($k$) for ($\beta < \alpha$) with parameters ($\beta = 0.5, \alpha = 0.7, \tau_1 = 5, \tau_2 = 11$) is shown in Fig. 3a. As expected (23), correlation leak around ($\tau_2 = 5$) results in statistically significant cross-correlation estimates at lags ($k = 10 \text{ and } 11$) considerably larger than those at ($\tau_1 = 5$). This in turn disrupts the ranking $R_{xy}(\tau_2) > R_{xy}(\tau_1) > R_{xy}(k)$. However, for $\beta > \alpha$ with ($\beta = 0.8, \alpha = 0.7$) dominant cross-correlation estimates occur at delays ($\tau_1 = 5, \tau_2 = 11$) preserving the ranking $R_{xy}(\tau_2) > R_{xy}(\tau_1) > R_{xy}(k)$, Fig. 3c. While the ranking is preserved for constraint $\beta > \alpha$, cross-correlation estimates at lags other than ($k = 5, 11$) corresponding to delays ($\tau_1$ and $\tau_2$) are rendered statistically significant. As seen earlier (Sec. B), these can indicate spurious existence of multiple delays between the driver and the dependent processes in addition to ($\tau_1$ and $\tau_2$). It is also important to note that the constraint $\beta > \alpha$ for preserving the rank turns out to be stringent especially in the limit $\alpha \to 1$ (5), i.e. the family of processes from which the delays can be inferred reduces dramatically as $\alpha \to 1$.

*Delay estimation from the increment process*

Cross-correlation between the increment series $\delta x_{n+1} = x_{n+1} - x_n$ and $\delta y_{n+1} = y_{n+1} - y_n$ delays $\tau_1$ and $\tau_2$ are given by

$$R_{\delta x \delta y}(\tau_1) = E(\delta x_n \delta y_{n+\tau_1}) = 2\beta.[R_x(0) - R_x(1)] + [2R_x(\tau) - R_x(\tau+1) - R_x(\tau-1)] \ \ldots\ldots \ (24)$$

$$R_{\delta x \delta y}(\tau_2) = E(\delta x_n \delta y_{n+\tau_2}) = 2.[R_x(0) - R_x(1)] + \beta[2R_x(\tau) - R_x(\tau+1) - R_x(\tau-1)] \ \ldots\ldots \ (25)$$

Substituting for the auto-correlation function $R_x()$ from (6) we get

$$R_{\delta x \delta y}(\tau_1) = [2\beta.(1-\alpha) - \alpha^{\tau-1}(1-\alpha)^2]R_x(0) \ \ldots\ldots\ldots\ldots\ldots\ldots\ldots\ldots\ldots\ldots\ldots\ldots. \ (26)$$

$$R_{\delta x \delta y}(\tau_2) = [2.(1-\alpha) - \beta\alpha^{\tau-1}(1-\alpha)^2]R_x(0) \ \ldots\ldots\ldots\ldots\ldots\ldots\ldots\ldots\ldots\ldots\ldots. \ (27)$$

It is important to note that the expressions (26) and (27) need not necessarily be positively correlated for every choice of the parameters ($\alpha$ and $\beta$). As noted earlier (Sec. B), we are interested in identifying only delays whose cross-correlation functions are positive. Therefore, prior to checking rank preservation $R(\tau_2) > R(\tau_1) > R(k)$, $\forall k \neq \tau_1, \tau_2$ we impose the constraint for positive cross-correlations at delays $\tau_1$ and $\tau_2$.

**Remark 3** *Constraint on parameters ($\alpha$ and $\beta$) such that $R_{\delta x \delta y}(\tau_1)$ and $R_{\delta x \delta y}(\tau_2)$ are positively correlated.*

Substituting for $R_{\delta x \delta y}(\tau_1)$ from (26) and imposing the constraint for positive correlation i.e. $R_{\delta x \delta y}(\tau_1) > 0$ we get

$$\beta > (1/2)\alpha^{\tau-1}(1-\alpha) \ldots\ldots\ldots\ldots\ldots\ldots\ldots\ldots\ldots\ldots \ldots\ldots\ldots\ldots\ldots \text{ (28)}$$

Subtracting (26) from (27) we get

$$R_{\delta x \delta y}(\tau_2) - R_{\delta x \delta y}(\tau_1) = (1-\beta).(1-\alpha)[2. + \alpha^{\tau-1}(1-\alpha)]R_x(0) \ldots\ldots\ldots\ldots\ldots \text{ (29)}$$

From (5) and (15) we know $0 < \alpha < 1$ and $0 < \beta < 1$, therefore

$$R_{\delta x \delta y}(\tau_2) - R_{\delta x \delta y}(\tau_1) > 0$$

From (28) and (29) we obtain

$$R_{\delta x \delta y}(\tau_2) > R_{\delta x \delta y}(\tau_1) > 0 \text{ for } \beta > (1/2)\alpha^{\tau-1}(1-\alpha). \ldots\ldots\ldots\ldots\ldots\ldots\ldots \text{ (30)}$$

While the constraint on the original processes (23) is a function of the parameter ($\alpha$), the constraint on the increment processes (30) is a function of the parameter ($\alpha$) as well as the differential delay ($\tau = \tau_2 - \tau_1$). It is important to note that the constraint on the increment process (30) is not as stringent as that on the original process (23) in general. For instance, cross-correlation analysis of the increment process, Figs. 3b and 3d, preserves the ranking $R_{xy}(\tau_2) > R_{xy}(\tau_1) > R_{xy}(k)$ for both the instances ($\beta > \alpha$) and ($\beta < \alpha$) discussed earlier, Figs. 3a and 3c. However, for the special case where the differential delay ($\tau = \tau_2 - \tau_1 = 1$), the constraint on $\beta$ for the increment process (30), ($\beta > \frac{1-\alpha}{2}$) can be considerably larger than those on the original process (23) $(\beta > \alpha)$ especially for $(\alpha < 1/3)$. Thus delay estimation on the original process as opposed to that of the increment process is preferred for $(\alpha < 1/3)$ and $(\tau = \tau_2 - \tau_1 = 1)$. An instance with parameters $(\alpha = 0.1, \beta = 0.3, \tau_1 = 10, \tau_2 = 11)$ is shown in Fig. 4a. For these choices of parameters constraint (23) is satisfied whereas constraint (30) is not. Therefore, the delays can be successfully estimated from the original processes, Fig. 4a and not from the increment processes, Fig. 4b. However, cross-correlation estimates of the original processes reveals delays $(\tau_1 = 9$ and $\tau_2 = 12)$ as being statistically significant in addition to $(\tau_1 = 10$ and $\tau_2 = 11)$, Fig. 4a. For $\tau > 1$, constraint (30) is considerably less stringent than constraint (23) irrespective of the choice of $\alpha$, encouraging estimation of the delay from increment process as opposed to the original process. An instance $(\beta = 0.05, \alpha = 0.7, \tau_1 = 10, \tau_1 = 12)$ where neither of the constraints (23 and 30) is satisfied is shown in Figs. 4c and 4d respectively. In such cases, it is not possible to estimate the delays using the techniques described in the present study.

Finally, we show in the following remark that the ranking of the cross-correlation $R(\tau_2) > R(\tau_1) > R(k) \forall k \neq \tau_1, \tau_2$ between the driver and the dependent processes is implicitly preserved in the increment series unlike those of the original series (23). Cross-correlation estimates satisfy $R_{\delta x \delta y}(\tau_2) > R_{\delta x \delta y}(\tau_1) > 0$ under constraint (30). The only possibility that can disrupt the ranking $R_{\delta x \delta y}(\tau_2) > R_{\delta x \delta y}(\tau_1) > R_{\delta x \delta y}(k), k \neq \tau_1, \tau_2$ is correlation leak around $\tau_1$ and $\tau_2$. In the following remark we show that correlation leak around $\tau_1$ and $\tau_2$ are strictly negative. Therefore, the only positive cross-correlations estimates on the increment processes occur at delays $\tau_1$ and $\tau_2$. i.e. $R_{\delta x \delta y}(\tau_2) > R_{\delta x \delta y}(\tau_1) > 0$ whereas $R_{\delta x \delta y}(k) < 0$ for $k \neq \tau_1, \tau_2$.

**Remark 4** $E(\delta \tilde{x}_{n+k} \delta \tilde{y}_{n+\tau_1}) < 0$ for any $k > 0, \tau > 0$

$E(\delta \tilde{x}_{n+k} \delta \tilde{y}_{n+\tau_1})$

$= \beta.[2.R_x(k) - R_x(k+1) - R_x(k-1)] + [2R_x(k+\tau) - R_x(k+\tau+1) - R_x(k+\tau-1)]$

Substituting for the auto-correlation function $R_x()$ from (5) we get

$$E(\delta \tilde{x}_{n+k} \delta \tilde{y}_{n+\tau_1}) = -(1-\alpha)^2 (\beta \alpha^{k-1} + \alpha^{k+\tau-1}) R_x(0) < 0 \ldots\ldots\ldots\ldots\ldots\ldots\ldots\ldots\ldots (31)$$

**Remark 5** $E(\delta \tilde{x}_{n+k} \delta \tilde{y}_{n+\tau_2}) < 0$ for $k > 0, \tau > 0$

Substituting for the auto-correlation function $R_x()$ from (5) we get

For $k > 0, \ 0 < \tau < k$

$$E(\delta \tilde{x}_{n+k} \delta \tilde{y}_{n+\tau_2}) = -(1-\alpha)^2 \alpha^{k-1} (\alpha^{-\tau} \beta + 1) R_x(0) < 0 \ldots\ldots\ldots\ldots\ldots\ldots\ldots (32)$$

For $k > 0, \ \tau > k$

$$E(\delta \tilde{x}_{n+k} \delta \tilde{y}_{n+\tau_2}) = -(1-\alpha)^2 (\alpha^{\tau-k-1} \beta + \alpha^{k-1}) < 0 \ldots\ldots\ldots\ldots\ldots\ldots\ldots\ldots (33)$$

**Summary II** *For a two-node network with two delays and Gauss-Markov driver ($0 < \alpha < 1$), delay estimation on the increment processes results in significant positive cross-correlation only at the respective delays $\tau_1$ and $\tau_2$ under constraint (30). This should be contrasted against delay estimation on the original processes where significant positive cross-correlations is observed at several lags in addition to that of $\tau_1$ and $\tau_2$.*

*Thus it is possible to identify multiple delays in addition to $\tau_1$ and $\tau_2$ on cross-correlation analysis of the original processes. Also, constraint (23) imposed on the original processes for preserving the rank $R(\tau_2) > R(\tau_1) > R(k)$, $\forall k \neq \tau_1, \tau_2$ is in general more stringent than the constraint (30) on the increment processes.*

### C. Long-range correlated driver with single and two-delays

Gauss-Markov driver process (5) considered in the above discussion is a short-range correlated driver whose correlation function decays exponentially as a function of lag (6). Non-markovian or long-range correlations have been observed in a wide-range of experimental systems [9-11] and accompanied by auto-correlation functions that decay as a power-law [9-11] with lag. Identifying delays from the original and increment processes for a two-node acyclic network with a long-range correlated driver is briefly discussed below.

<u>Power-law correlated driver</u>

Auto-correlation function of classical long-range correlated noise exhibit power-law decay at large time scales ($k$) and follows the generic form [9, 10].

$$R_x(k) = k^{-\gamma}, \text{ where the Hurst exponent } \gamma \text{ lies in the interval } (0.5,1) \ldots\ldots (34)$$

The auto-correlation function (34) is positive and decays monotonically as function of the lag $k$.

C1. <u>Long-range correlated driver and single delay</u>

Consider the driver process (34) and the dependent process (Sec. B1)

$$y_n = x_{n-\tau} \ \ldots\ldots\ldots\ldots\ldots\ldots\ldots\ldots\ldots\ldots\ldots\ldots\ldots\ldots\ldots\ldots.\ldots\ldots (35)$$

*Delay estimation from the original process*

Following procedure similar to (Sec. B1) we get

$$E(x_{n+k} y_{n+\tau}) = R_x(k) > 0 \ \forall \text{k} \ \dots\dots\dots\dots\dots\dots\dots\dots\dots\dots\dots\dots\dots \text{(36)}$$

Also from (34)   $R_x(k) < R_x(0)$ $\dots\dots\dots\dots\dots\dots\dots\dots\dots\dots\dots\dots\dots\dots\dots$ (37)

As in the case of Gauss-Markov process (9, 10, Sec. B1) positive cross-correlations persist for lags other than delay $\tau$.

*Delay estimation from the increment process*

Following procedure similar to (Sec. B2) we get

$$E(\delta x_n \delta y_{n+\tau}) = 2[R_x(0) - R_x(1)] > 0 \dots\dots\dots\dots\dots\dots\dots\dots\dots\dots \text{(38)}$$

$$E(\delta x_{n+k} \delta y_{n+\tau}) = 2R_x(k) - R_x(k+1) - R_x(k-1) \dots\dots\dots\dots\dots\dots\dots\dots \text{(39)}$$

Substituting for $R_x(k)$ from (34) into (39) we get

$$E(\delta x_{n+k} \delta y_{n+\tau}) = k^{-\gamma}[2 - (1 + \frac{1}{k})^{-\gamma} - (1 + \frac{1}{k})^{-\gamma}] \dots\dots\dots\dots\dots\dots\dots\dots \text{(40)}$$

Binomial expansion of (40) under the assumptions in (34), i.e. $k >> \gamma$ and $0 < \beta < 1$ we get

$$E(\delta x_{n+k} \delta y_{n+\tau}) = -k^{-\gamma}[\frac{\gamma(\gamma+1)}{k^2} + .....] < 0 \dots\dots\dots\dots\dots\dots\dots\dots\dots\dots \text{(41)}$$

The above expression (41) is negative for $\forall k \neq 0$. As in the case of Gauss-Markov driver (12, 13, Sec. B1) $E(\delta x_n \delta y_{n+\tau}) > 0$ whereas $E(\delta x_{n+k} \delta y_{n+\tau}) < 0$ for $\forall k \neq 0$.

## C2. Long-range correlated driver and two delays

Consider the case of two delays, where the driver process $x_n$ satisfies (34) and the dependent process (Sec. B2) satisfies

$$y_n = \beta . x_{n-\tau_1} + x_{n-\tau_2} ; 0 < \beta < 1, \tau_2 > \tau_1 > 0 \ldots\ldots\ldots\ldots\ldots\ldots\ldots\ldots.. \ (42)$$

*Delay estimation from the original process*

As in the case of the Gauss-Markov process (20) we obtain

$$R_{xy}(\tau_2) > R_{xy}(\tau_1)$$

Constraint on the parameter ($\beta$) (23) in order to preserve the ranking $R(\tau_2) > R(\tau_1) > R(k), \forall k \neq \tau_1, \tau_2$ is

$$\beta > \frac{R_x(1) - R_x(\tau)}{(R_x(0) - R_x(\tau-1))} \ldots\ldots\ldots\ldots\ldots\ldots\ldots\ldots\ldots\ldots\ldots\ldots\ldots\ldots\ldots. \ (43)$$

*Delay estimation from the increment process*

Following procedure similar to (Sec. B2) and from the binomial expansion (41) it is possible to obtain a constraint for $R_{\delta x \delta y}(\tau_2) > R_{\delta x \delta y}(\tau_1) > 0$. Following procedure similar to Remarks 4 and 5 and using the binomial expansion (41) it can be shows that $E(\delta x_{n+k} \delta y_{n+\tau_1}) < 0$ and $E(\delta x_{n+k} \delta y_{n+\tau_2}) < 0$.

**Summary III** *As in the case of Gauss-Markov process (Summary I and Ii), delay estimation on the increment process of long-range correlated driver can significantly minimize the impact of spurious identification of delays between the driver and the dependent process.*

An instance of delay estimation from two-node acyclic network with long-range correlated driver and with one and two-delays is shown in Figs. 5 and 6. Long-range correlated driver process was generated from stationary fractional auto-regressive integrated moving average process FARIMA (0, d, 0) with Gaussian innovations and parameter $d = 0.3$ [9, 10]. This corresponds to Hurst exponent $\gamma = d + 0.5$ (34). Cross-correlation analysis $R_{xy}(k)$ between FARIMA (0, d, 0) driver and the dependent process ($y_n = x_{n-\tau}$, $\tau = 10$, N = 4000) along with those of their increment processes $R_{\delta x \delta y}(k)$ is shown in Figs. 5c and 5d respectively. As seen earlier, delay estimation of the increment process minimizes spurious statistically significant delays. Cross-correlation estimates for long-range correlated driver and dependent process $y_n = \beta.x_{n-\tau_1} + x_{n-\tau_2}$, with parameters ($\beta = 0.5, \tau_1 = 5, \tau_2 = 11, N = 4000$) is shown in Figs. 6c and 6d respectively. The ranking $R_{\delta x \delta y}(\tau_2) > R_{\delta x \delta y}(\tau_1) > R_{\delta x \delta y}(k), k \neq \tau_1, \tau_2$ is preserved on cross-correlation analysis of the increment process, Fig. 6d. This has to be contrasted to analysis of the original process where the ranking is not preserved, Fig. 6c. Analysis of the original process also reveals statistically significant cross-correlation estimates at several lags in addition to ($\tau_1 = 5$ and $\tau_2 = 11$).

**D. Delay estimation from coarse-grained realizations**

Coarse-grained realizations are simplified representations of the actual processes. An example is that of a one-dimensional ising spin model where each element is either an up (+1) or a down (-1) spin. In the present study, we generate coarse-grained realizations of the given process about their mean, E(*x*), given by

$$x_c^i = +1 \text{ if } x_i > E(x)$$
$$\qquad = -1 \text{ otherwise}$$ ……………………………………………………….. (44)

For stationary zero-mean normally distributed processes, an analytical expression can be derived relating the correlation of the original process $R_x(k)$ to that of its coarse-grained counterpart $R_{xc}(k)$ [12, 13], given by

$$R_{x_C}(k) = \frac{2}{\pi} \arcsin \frac{R_x(k)}{R_x(0)}$$ …………………………………………………... (45)

It is important to note that in Sec. B and C, the short-range (5) and the long-range (34) correlated driver were generated as linear combinations of normally distributed variables, hence normally distributed. This in turn implies that coarse-grained realizations about the mean of the driver processes (5 and 34) follow relation (45). Since the short-range and long-range driver processes $R_x(k)$ considered have monotonic decreasing auto-correlation function, those of their coarse-grained counterpart $R_{x_C}(k)$ (45) are also monotonic decreasing. Dependent processes $y_n = x_{n-\tau}$ and $y_n = \beta.x_{n-\tau_1} + x_{n-\tau_2}$ of the normally distributed driver $x_n$ are linear combinations of normal processes, hence implicitly normal. Thus coarse-grained representation of the driver process and the dependent processes about their means follows relation (45). It should also be noted that the corresponding increment series by definition is the difference of normally distributed processes, hence normal.

In the following discussion, coarse-grained realizations of the original and the increment driver $x_n$ and dependent $y_n$ processes shall be represented by $x_{n_C}$ and $y_{n_C}$ respectively. The coarse-grained realizations of the increment processes ($\delta x_n$ and $\delta y_n$) are represented by ($\delta \hat{x}_{n_C}$ and $\delta \hat{y}_{n_C}$). The cross-correlation estimates on the coarse-grained original and increment processes are represented by $R_{x_C y_C}(k)$ and $R_{\delta \hat{x}_C \delta \hat{y}_C}(k)$.

We show instances where $R_{\delta \hat{x}_C \delta \hat{y}_C}(k)$ is useful identifying delays whereas unlike $R_{x_C y_C}(k)$. This is demonstrated on the two-node acyclic networks with short-range and long-range correlated drivers with one and two-delays. Cross-correlation estimates $R_{x_C y_C}(k)$ for the coarse-grained realizations of the Gauss-Markov driver (5) with ($\alpha = 0.9$, N = 4000) and the dependent process ($y_n = x_{n-\tau}$, $\tau = 10$), along with those of their increment series $R_{\delta \hat{x}_C \delta \hat{y}_C}(k)$ is shown in Figs. 5e and 5f respectively. Cross-correlation estimates $R_{xy}(k)$ and $R_{\delta \hat{x} \delta \hat{y}}(k)$ obtained on $x_n$ and $y_n$ is shown in Figs. 5a and 5b for qualitative comparison. It is important to note that the estimation on the increment series results is minimizing the effect of correlation leak as observed earlier (Summary I). Similar results were obtained in the case of FARIMA (0, d, 0) driver with ($d = 0.3$, N = 4000), Figs. 5g and 5h. These results conform to earlier observations (Summary I and III), where analysis of the increment processes minimize statistically significant false-positive correlation.

Cross-correlation estimates $R_{x_C y_C}(k)$ for the coarse-grained realizations of the Gauss-Markov driver (5) with ($\alpha = 0.9$, N = 4000) with dependent process ($y_n = \beta.x_{n-\tau_1} + x_{n-\tau_2}$, $\tau_1 = 5$, $\tau_2 = 11$, N = 4000) is shown in Fig. 6e. Cross-correlation analysis of coarse-grained counterparts of the corresponding increment series $R_{\delta \hat{x}_C \delta \hat{y}_C}(k)$ is shown in Fig. 6f. A similar analysis of the FARIMA (0, d, 0) driver with ($d = 0.3$, N =

4000) is shown in Figs. 6g and 6h. These results conform to earlier observations (Summary II and III), where analysis of the increment processes minimizes statistically significant correlation leak and preserves the rank ordering $R(\tau_2) > R(\tau_1) > R(k)$, $k \neq \tau_1, \tau_2$. Therefore, analysis of the increment process can minimize statistically significant false-positive correlations even in the case of coarse-grained counterparts.

## 3. Discussion

The present study, investigated statistical estimation of delays between the driver and dependent processes in a two-node acyclic network with one and two delays using linear measures such as cross-correlation function. While delay estimation is straightforward in the case of uncorrelated drivers, correlated drivers can result in significant correlation leak around the actual delay between the driver and dependent process. Such correlation leak can result in spurious identification of statistically significant delays and existence of multiple paths between the driver and dependent process. Cross-correlation analysis of the increment processes was shown to significantly minimize the effect of correlation leak under certain constraints. In the presence of two-delays between the driver and the dependent node, cross-correlation analysis of the increment processes preserved the ranking of the auto-correlation function in addition to identifying the delays. This was demonstrated on short-range correlated Gauss-Markov process whose auto-correlation function decays exponentially and long-range correlated FARIMA (0, d, 0) driver with power-law decaying auto-correlation function. Correlation properties of stationary normal processes are analytically related to correlation of their corresponding coarse-grained counterpart generated about their mean. An instance was shown where cross-correlation estimates on the coarse-grained realizations of the increment series significantly minimized the effect of correlation leak. Thus from the above results cross-correlation analysis of the increment processes can provide insight into the nature of delays not evident from the analysis of the original processes.

## 4. Acknowledgement

The present study was supported by funds from National Library of Medicine (1R03LM008853-1) and junior faculty grant from American Federation for Aging Research (AFAR).

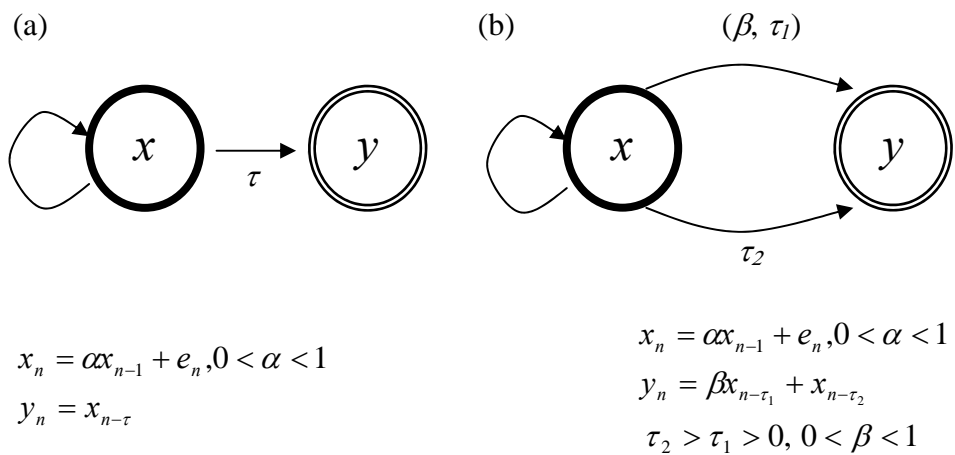

**Figure 1** Two-node acyclic networks with one and two-delays are shown in (a) and (b) respectively. The driver and the dependent processes are represented by ($x$) and ($y$).

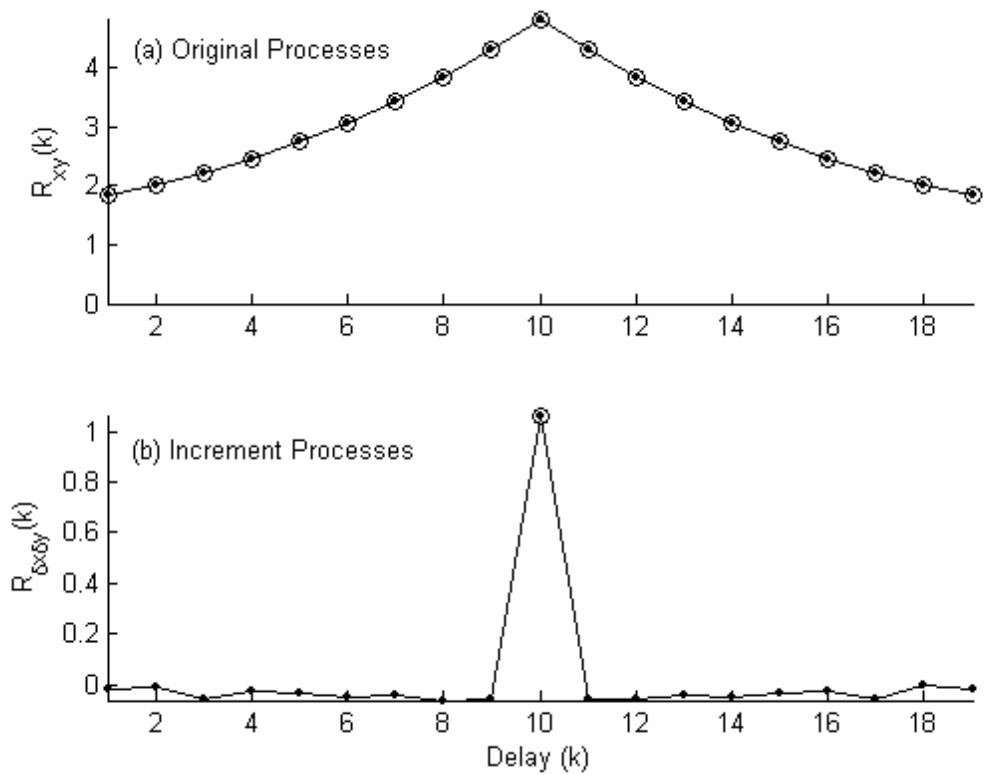

**Figure 2** Cross-correlation estimates as a function of delay ($k$) for the original $R_{xy}(k)$ (a) and increment processes $R_{\delta x \delta y}(k)$ (b) in a two-node acyclic network with a single delay ($\tau = 10$) and Gauss-Markov driver ($\alpha = 0.9$, N = 4000). Statistically significant delay estimates ($n_s = 99$, Sec A) are shown by circles.

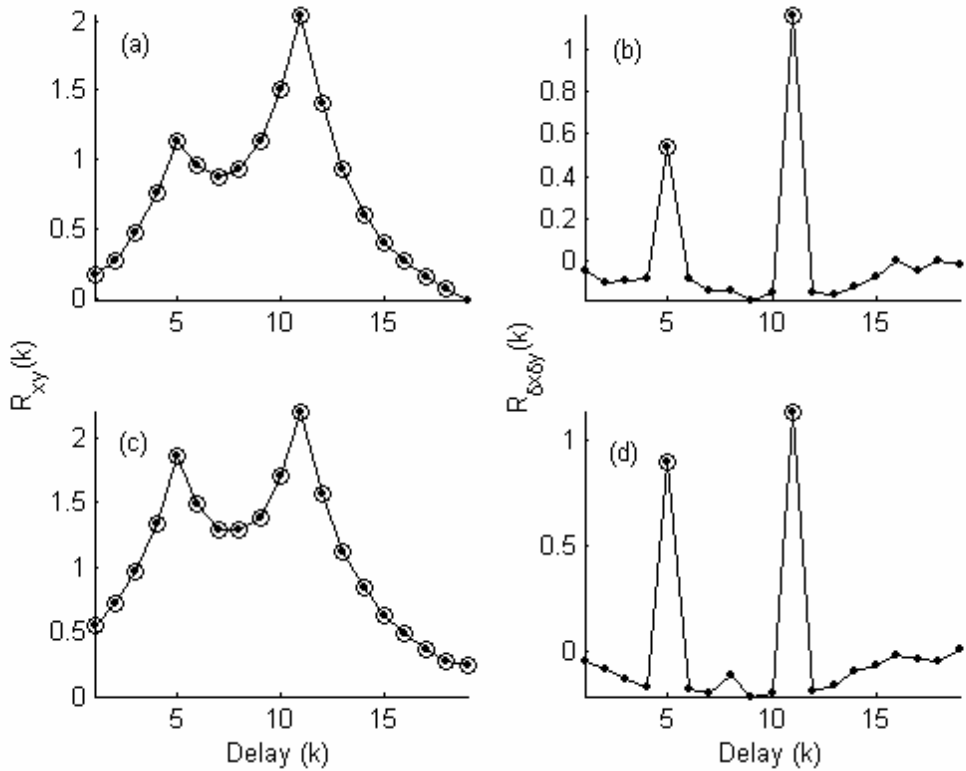

**Figure 3** Cross-correlation estimates as a function of lag ($k$) of the original $R_{xy}(k)$ (a, c) and increment processes $R_{\delta x \delta y}(k)$ (b, d) in a two-node acyclic network with two delays ($\tau_1 = 5, \tau_2 = 11$) and Gauss-Markov driver ($\alpha = 0.7$, N = 4000). Cross-correlation estimates $R_{xy}(k)$ for original processes violating constraint (23) i.e. ($\beta < \alpha, \beta = 0.5, \alpha = 0.7$) is shown in Fig 3a. Those of its increments $R_{\delta x \delta y}(k)$ are shown in Fig. 3b. Cross-correlation estimates of original processes $R_{xy}(k)$ satisfying constraint ($\beta > \alpha, \beta = 0.8, \alpha = 0.7$) is shown in Figs. 3c. Those of its increments $R_{\delta x \delta y}(k)$ are shown in Fig. 3d. Statistically significant delay estimates ($n_s = 99$, Sec. A) are shown by circles.

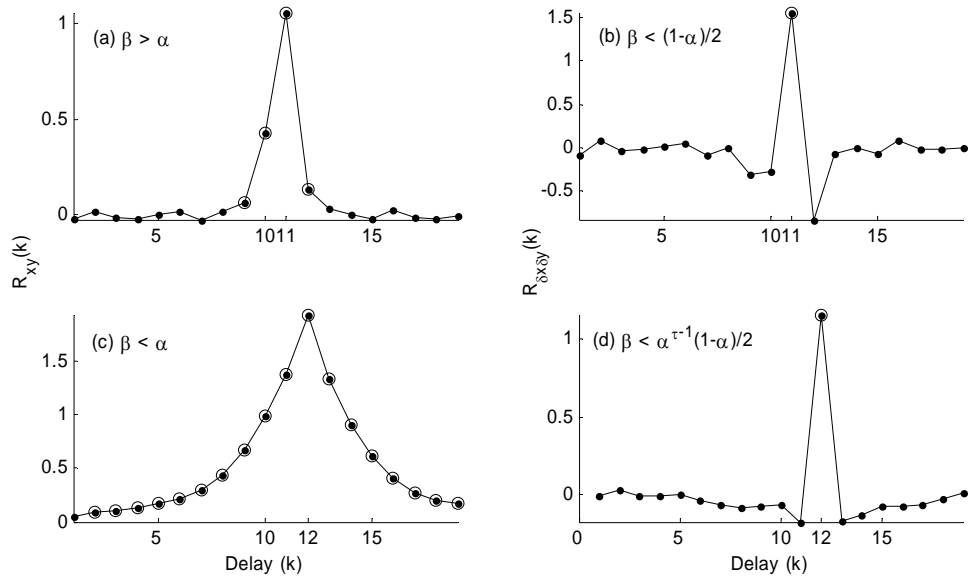

**Figure 4** Cross-correlation estimates as a function of lag ($k$) of the original $R_{xy}(k)$ (a) and increment processes $R_{\delta x \delta y}(k)$ (b) in a two-node acyclic network with two delays ($\tau_1 = 10, \tau_2 = 11$) and Gauss-Markov driver ($\alpha = 0.1$, N = 4000). Cross-correlation estimates that satisfy constraints (23) and (30) i.e. ($\beta = 0.3, \alpha = 0.1, \tau = 1$) for the original and increment processes is shown in Figs. 4a and 4b respectively. Cross-correlation estimates for the original and increment processes with parameters ($\beta = 0.05, \alpha = 0.7, \tau_1 = 10, \tau_1 = 12$) is shown in Figs. 4c and 4d respectively. Statistically significant delay estimates ($n_s = 99$, Sec. A) are show by circles.

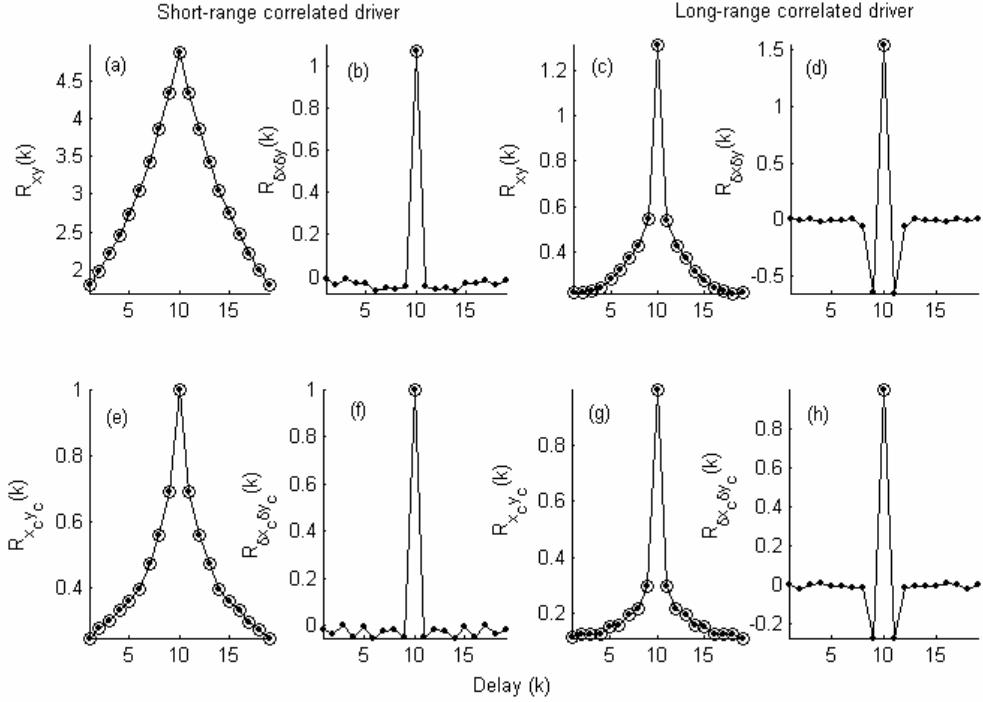

**Figure 5** Cross-correlation estimates as a function of lag ($k$) for the original $R_{xy}(k)$ (a) and increment processes $R_{\delta x \delta y}(k)$ (b) in a two-node acyclic network with a single delay ($\tau = 10$), with Gauss-Markov ($\alpha = 0.9$, N = 4000) (a, b) and FARIMA (0, d, 0) driver (d = 0.3, N = 4000) (c, d). Cross-correlation estimates of the corresponding coarse-grained realizations of the original $R_{x_C y_C}(k)$ and increment series $R_{\delta x_C \delta y_C}(k)$ of the Gauss-Markov (e, f) and FARIMA(0, d, 0) (g, h) driver is shown right below them. Statistically significant delay estimates ($n_s = 99$, Sec. A) are show by circles.

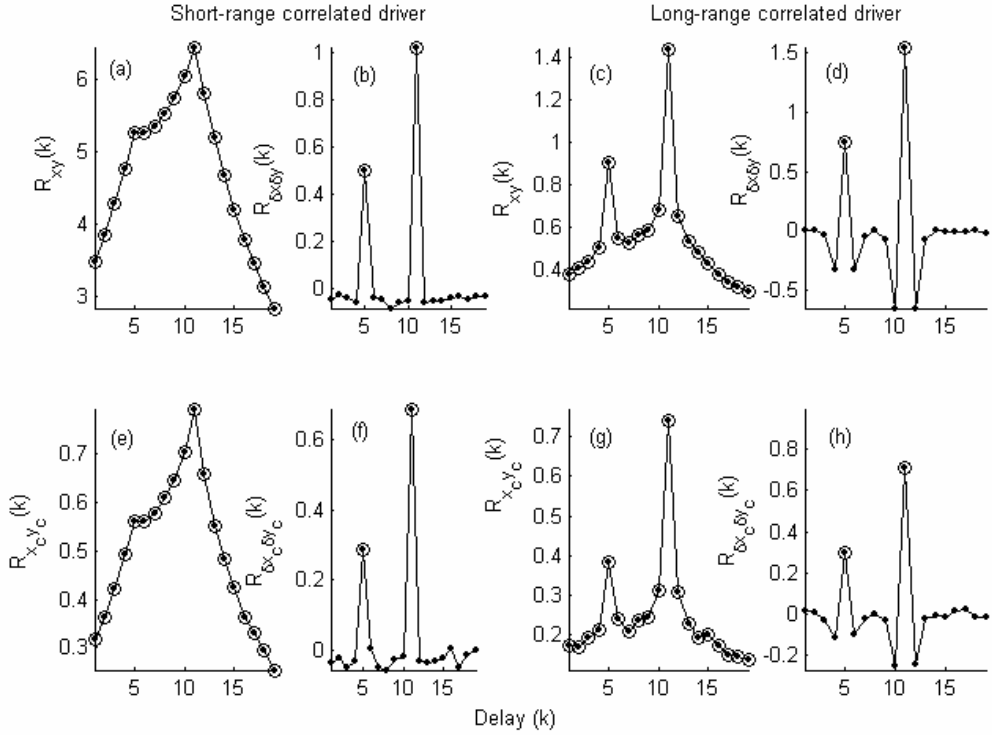

**Figure 6** Cross-correlation estimates as a function of lag ($k$) for the original $R_{xy}(k)$ (a) and increment processes $R_{\delta x \delta y}(k)$ (b) in a two-node acyclic network with two delays ($\tau_1 = 5, \tau_2 = 11, \beta = 0.5$) with Gauss-Markov ($\alpha = 0.9$, N = 4000) (a, b) and FARIMA (0, d, 0) driver (d = 0.3, N = 4000) (c, d). Cross-correlation estimates of the corresponding coarse-grained realizations of the original $R_{x_C y_C}(k)$ and $R_{\delta x_C \delta y_C}(k)$ increment series corresponding to the Gauss-Markov (e, f) and FARIMA(0, d, 0) (g, h) driver is shown right below them. Statistically significant delay estimates ($n_s = 99$, Sec. A) are show by circles.